# SCRUTINY NEW FRAMEWORK IN INTEGRATED DISTRIBUTED RELIABLE SYSTEMS

Mehdi ZekriyapanahGashti

Department of Computer Engineering
Payame Noor University
I.R of IRAN
Zekriya@pnu.ac.ir

*ABSTRACT*

*In this paper we represent a new framework for integrated distributed systems. In the proposed framework we have used three parts to increase Satisfaction and Performance of this framework. At first we analyse integrated systems and their evolution process and also ERPSD and ERPDRT framework briefly then we explain the new FDIRS framework. Finally we compare the results of simulation of the new framework with presented frameworks .Result showed In FIDRS framework, the technique of heterogeneous distributed data base is used to improve Performance and speed in responding to users. Finally by using FDIRS framework we succeeded to increase Efficiency, Performance and reliability of integrated systems and remove some of previous framework's problems.*

## 1. INTRODUCTION

In recent years there was a tendency toward analysing integrated systems especially about new framework for these systems. Integrated systems integrate all departments and tasks across the resources in an integrated system. That are supported the specific needs of all parts of the system. Considering software this is a complex subject that a computer program can support financial part and human resources in addition to storage part and other parts of the organization. Each of these sections according to their tasks has an independent system but integrated system collects all these subsystems in a single system for updating and distributing data and information between different parts easily.Information has become an organization's most precious asset. Organizations have become increasingly dependent on information [1]. So communication between departments will be easier and agencies will get significant advantages[2].

A distributed computing system or parallel systems is defined as the collection of computers (either homogenous or heterogeneous) or workstations [3]. Resources integrated system, old autonomous finance, resources, manufacturing and storage systems has replaced with an integrated system of software modules. Financial parts, production parts and storage parts do a special task in theirs modules but internal and operational link between these modules causes each person of finances part can access the storage system and status of activities related to a specific order. Some manufacturers of integrated systems are provided modules independently of each other without the need to purchase and install the whole system. In this case, the other modules can be obtained at the following times [4]. Companies can make the modules based on their ability and then integrate them.

Integrated systems are different considering quality rather than of formal methods because organization structure connects business processes and information technology systems in a unified framework. Integrated systems is an innovative technology that have many benefits such as, the speed of decision, making cost reduction and control of all trade and service by managers. Organization must be Process Oriented to implement integrated systems.

Now day solutions of information technology are growing in organizations. When every unit of organization starts its task to separately, so the general objectives of the organization are forgotten and senior manager can't analyse various reports of functions and duties of different parts and can't decide based on it in proper time. Hence integrated systems technology is created to automate all processes in each unit without cable or wireless connections [5]. Resource Integrated System is not just a program. It is a collection of thoughts, architecture, performance, and motivation to achieve the goals of an organization and stratification of customers. Resource Integrated System is expanded from an idea for producing to a comprehensive solution for all activities and services.

In recent decades, Satisfaction levels of customers and users of integrated systems along to increasing orders for faster service, with a wide range of services at lower prices has changed. So many modern information systems use new technologies like SCM or RFID to identify the goods and services. Radio Frequency Identification (RFID) technology is one of the most important technologies in this decade. This technology allows identifying the tagging objectives wirelessly using transponders queried by readers through a wireless channel. RFID technology has widely been used in applications such as public transportation, supply chain management, e-passports, location tracking systems and access control systems [6]. Supply chain management applications in production industries and their supply chains consist of very complex techniques[7].

In the beginnings of creating integrated system, a wide number of organizations accept it enthusiastically and some others reject it, now day's integrated systems can grow on different science such as the science about aerospace, defence, automotive industry and health. Also There is a new performance opportunities for all classes to update and add new users as well [8]. In recent decades, Oracle, SAP, PeopleSoft and SSA Baan Company had about 70% of market share in the production of integrated Systems (See figure 1).

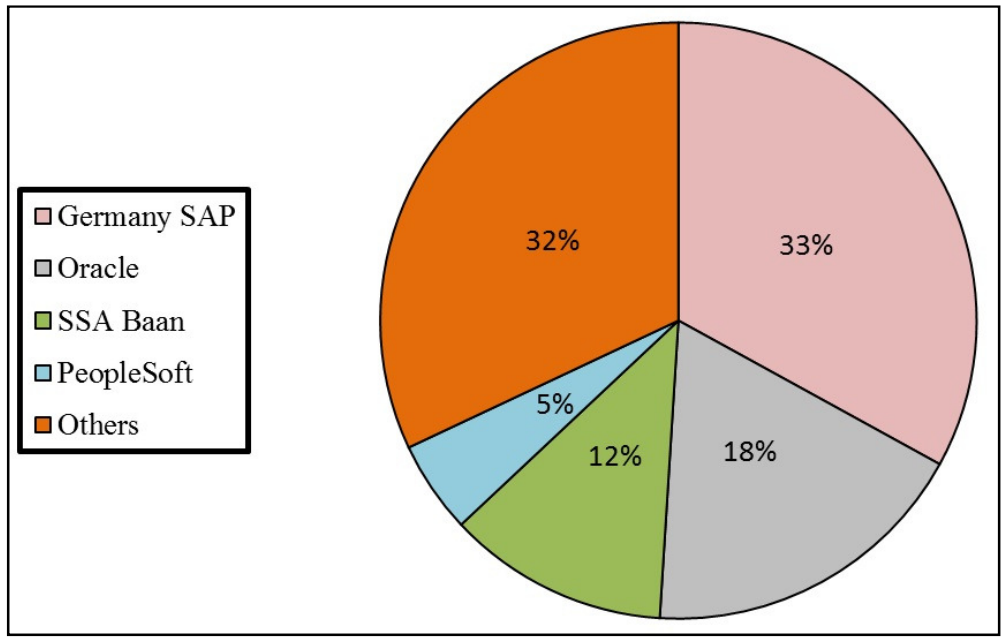

Figure1. Shares of producer of integrated systems [8]

## 2. A REVIEW OF PAST FRAMEWORKS

Integrated systems are findings of IT that are used in a lot of countries around world and are changing and developing rapidly. Although integrated systems have been borne since two decadesago, but they can be considered historically the continuance of a movement that has started from systems of material requirement planning and has developed gradually. Organizational resource planning systems are borne because of funds competitions and their abilities. Planning organizational resources is the attempt in integrating all parts and operations of an organization in a general electronic system that is able to provide all special needs of different parts.

As we know, data changes to information after processing and when this information, has a specific order and arrangement changes into knowledge that in this case comes closer to technology. But the point is that data should possess higher security in order to make a possible for us to use it reliably and securely in the electronic world. These issues have been analysed in details in integrated systems and by integrating them we will be able to increase practicality and beneficiary in a complete information system. Following there is a description of some frameworks for integrated systems

### 2.1. ERPSD Framework

One of parameters of this method in integrated software packages is its proficiency that is explained in terms of index codes of large information banks. So in this way we can gain its proficiency that is explained in terms of index codes of large information banks. So in this way we can gain its complete features and characteristic by gathering elements information and using inside and outside networks. It can have influences on old systems because old systems were not based on index code but instead were defined as sub index and availability of information was concentrated so proficiency was low in old systems but authors have succeeded to increase proficiency by help of EPRSD packages that has access direct to inside and outside entities , It means that a user can reach all features of an object from elementary threshold to secondary threshold having the general index code , so in comparison to previous versions proficiency has increased and also quick access to information becomes possible by ERPSD sub system. This subsystem is a distributed system .In a distributed system each level but of course each level needs service, too and the lower level serves it [9].

ERPSD uses distributed data base that is in agreement with Internet and because of this agreement is more reliable that its previous versions, and as its central data is distributed and from every regions of inside and outside net can connect with information bank and transform data more rapidly. We should pay attention to this point that a distributed system costs more than a central system.

In ERPSD it has been used of sharing sources in inside nets to decrease cost. In traditional system it is used of firewall to increase security but in an ERPSD we use a distributed firewall in other words by changing subsystem there won't be any general changes to the general framework of ERPSD because the firewall is distributed in ERPSD[10].

### 2.2. ERPDRT Framework

ERPDRT framework in integrated systems helps to increase reliability and the ability to be distributed in the system by using simultaneity technique. In ERPDRT different levels of availability are used to increase customer's satisfaction. Inner, outer and middle levels in establishing simultaneity are among different phases of relationship management with customers. Availability level helps us to increase the security of ERPDRT framework in comparison with previous frameworks. When a customer refers to integrated distributed systems which are reliable too, or if they are among customers whom individual information has been registered before some conditions will appear. Condition A is when there is no information about admissive customer in different parts of system so can't have simultaneous access In this condition we can consider more time for system in order to register customers ' individual information and then system can stay in one of following conditions. Condition B is when customer's individual information has been registered in information bank. We can represent a more reliable and proficient service to customers [11].

In ERPDRT framework it has been used of ERPASD algorithm. ERPASD made a new technique which decreased repetitive data in proceeding .For simulating ERPASD algorithm in ERPDRT framework data which was used was from Ministry of education. After simulating the algorithm and checking time of doing the Apriori algorithm and ERPASD algorithm some results were found which showed positive effects during simulating the proposed algorithm and ERPDRT framework and the ERPASD algorithm was more effective than Apriori algorithm. The basic Apriori algorithm had some limitations in searching information banks but ERPASD algorithm presented a new technique which decreased problems more than previous frameworks and by the help of this algorithm , information processing way improved . This work possessed higher proficiency and reliability than its previous versions [12].

## 3. THE NEW FIDRS FRAMEWORK

After studying about previous frameworks and doing simulating them in different companies and organizations Some problems found, so we're commended a new framework called FIDRS to decrease and remove problems [13]. We used RMSD algorithm in FIDRS framework because of our customers satisfaction and improving our services to them, also we used compound data basis in searching phase of information banks. Following is a description of all parts of FIDRS framework.

### 3.1. Customers Relationship Management phase (CRM)

CRM is a complete way for identifying, attracting and holding customers. Also CRM enables organizations to manage and harmonize relationships with customers through some channels, parts, commercial and geographical ways. CRM is a commercial way that other people, processes and technology to maximize organization's relationships with customers. The true importance of CRM is changing the strategy, practical processes and commercial, business performances in order to attract and hold customers and increase productivity. CRM is a strategy which its goal is understanding predicting and managing organizations and customer's needs.

#### 3.1.1. Information Resource Management (IRM)

Organization's information is not only organization's property source but also is a tool for managing other sources and properties of organization. This Value is not practical unless necessary information will be in reach of authorized person in an appropriate time. If data explains summary of relationships between facts, Information will be the definitions which are attributed to data and make a bigger set. IRM is one of tools which defines and describes the process of work and information flow in organs.

#### 3.1.2. Sell Configuration & Services System (SCSS)

Often it is believed that sell part and service part should work well together to help organization to improve its conditions. Sell part always regrets that customer service part mentions just minor problems and customer service part believes that sell part throws organization in trouble by giving false and unreal promises. Don't you think that organizations sell their products by these unreal promises? Truth is that for winning in this game sell and service part must play in the same team. Sell configuration and services system is responsible for making a balance between these two parts. This system considers some exceptions for customers to make them satisfied.

### 3.1.3. Strategy

Strategy is a macro program for reaching a unique aim. This term originates from wars planning. CRM strategy defines itself as a major program for gaining the goal protecting and improving it in an organization. Each organization in the deal world should have a strategy for CRM. The most important factor in success of different organizations is customer's satisfaction. Balanced scorecard (BSC) is one of tools for analysing according to financial norms that organizations use them for studying and estimating customer's satisfaction. If necessary information about customers doesn't exist or it is seldom, we can make two concise groups for customers before performing CRM strategy. If in an organization there is such a condition, before performing CRM strategy it should be a study on customers satisfaction phase.

### 3.1.4. Risk Management

There are different definitions for risk management that all of them have the same meaning and are focused on risk management process such as: Risk management is the process of recognizing, decreasing them down to a reasonable level and finally analysing its results on the system. Williams and Heinz define risk management so: "Risk management is the process of recognizing, analysing, and controlling accidental risks that its possible results are damage or unchanging in the condition. Risk management manages risks by controlling them and providing financial damages which have happened beyond efforts for damage control [14]. The most important goal of risk management is to help organization to manage risks better, and the goal of CRM risk management is managing risks which are related to CRM missions such as holding a continuous relationship with organization's customers.

## 3.2. Data Base

Data base is a set of data which is related to minimum unnecessary and extended applications that are independent of electronic and hardware programs. Database is organized in a special way in order to be able to retrieve data if it is needed. They use heterogeneous distributed data base system HDDBS in FIDRS to make an easier relationship among the new framework phases. In this database they use a standard entitled Gateway Protocol or the same APIs which connect DBMS with Applications which are used to connect different DBMSs such as ODBC, JDBC, etc.

## 3.3. Decision Support System phase (DSS)

Decision making system is an information system which has been designed to help managers in decision making. This system uses data models to solve complex management issues the .major goal in this system is informing manager's about the standard information about their companies and the outside world. This information should consist of time history of processes and organization's outcomes in order to cast future.

### 3.3.1. Database Management system (DBMS)

Database has been established by DBMS, and it's up to date by the help of DBMS. Establishing and controlling a database is very hard and complex. DBMS software makes equal all vast and complex files by using a special technique. The recommended framework consists of several databases which some of them are outside of organization. We must add manager's files to database, too DBMS responsible for managing all of these databases.

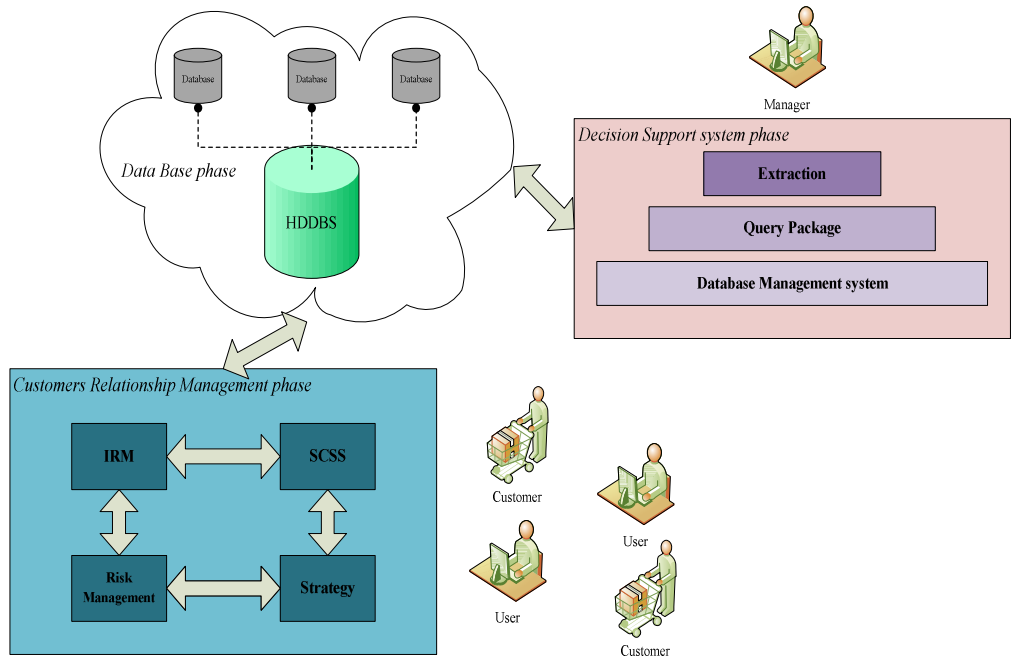

Figure 2. FIDRS Framework

### 3.3.2. Query Package

These packages consist of statistical and mathematical models such as linear planning and regression analysing or special programs for managers or an organization or even an industry. Noticeable models in this field are: strategic and ling run programming, tactical and operational programming and financial programming models.

### 3.3.3. Extraction

We need a survey or measurement layer for using data and rules in data base. In this layer we used RMSD algorithm to help us to optimize time complexity waiting time and surveying source timing by the help of a new algorithm [15].

## 4. SIMULATION FIDRS FRAMEWORK

For simulation the responding time to users and customer's requests in FIDRS, ERPSD and ERPDRT framework it was used of a set of software and hardware equipment's which has been named in table 1.

Table 1. Hardware and Software used for Simulation

| Hardware or Software | Information |
| --- | --- |
| Operating system | Microsoft Windows 7 Professional Service Pack 1 |
| Architecture | 64-bit Operating System |
| Processor | Intel 3.2 GHz |
| Memory (RAM) | 4 GB |

The results of comparing responding times which were gained in previous framework and FIDRS framework and have been presented in table 2.

Table 2. Result respond time of simulation FIDRS framework

| Number of Request | Respond time in ERPSD (Second) | Respond time in ERPDRT (Second) | Respond time in FIDRS (Second) |
|---|---|---|---|
| 50000 | 0.052 | 0.030 | 0.029 |
| 500000 | 0.803 | 0.599 | 0.503 |
| 3000000 | 2.031 | 1.382 | 1.057 |
| 5000000 | 3.109 | 2.178 | 1.996 |

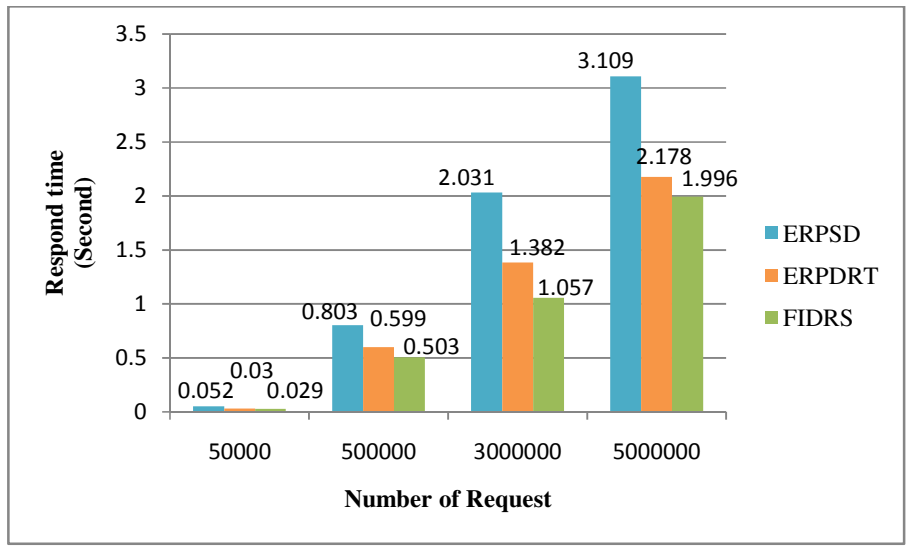

Figure 3. Compared total new FIDRS framework and past framework

## 5. CONCLUSIONS

In this paper we studied a new framework for integrated systems which a distributed technology was used in it. Results showed that in simulation of FDRS framework and comparing it with other frameworks that this framework is about 0.03% more practical than other frameworks in some conditions with very low number of requests. The increase of number of requests causes to improve effectiveness present and performance of FIDRS 15% and 8.7 % as compared with ERPSD and ERPDRT respectively.

International Journal of Distributed and Parallel Systems (IJDPS) Vol.3, No.5, September 2012
[6] M.H.Habibi, M.Gardeshi, M. Alaghband ,"Cryptanalysisof two mutual authentication protocols for low-cost RFID" International Journal of Distributed and Parallel Systems (IJDPS) Vol.2, No.1, January 2011.

[7] Manik Sharma, Gurdev Singh ,HarsimranKaur," A STUDY OF BNP PARALLEL TASK SCHEDULING ALGORITHMS METRIC'S FOR DISTRIBUTED DATABASE SYSTEM", International Journal of Distributed and Parallel Systems (IJDPS), January 2012

[8] M. Markus, C. Tanis, and P. Fenema, "Multisite ERP implementation,"Communications of the ACM, vol. 43, PP 42-46, 2004.

[9] A.GhorbanniaDelavar "ERPSD: A New Model for Developing Distributed, Secure, and Dependable Organizational Softwares" (IEEE ,CSIT' 2009 September 28-2 oct 2009

[10] A.GhorbanniaDelavar , L.Saiedmoshir " Analysis of a New Model for Developing Concurrent Organizational Behaviors with Distributed, Secure and DependableERPSDTechnology " (IEEE, CSIT 2007 September 24-28 2007 , PP 321-324)

[11] ArashGhorbanniaDelavar, BehrouzNoori, Mehdi ZekriyapanahGashti , "ERPDRT : A Novel Real-time Framework in Integrated Distributed System Resources, Secure with Data Mining Mechanisms", IEEE ICCSN 2011, May 2011, Xi'an, China

[12] ArashGhorbanniaDelavar , Mehdi ZekriyapanahGashti, BehrouzNooriLahrood, "ERPASD: A Novel Algorithm for Integrated Distributed Reliable Systems Using Data Mining Mechanisms ", IEEE ICIFE 2010, September 2010, Chongqing China

[13] Mehdi ZekriyapanahGashti, " FIDRS: A Novel Framework for Integrated Distributed Reliable Systems", IJCSI International Journal of Computer Science Issues, Sept 2012

[14] Gary Stoneburner, Alice Goguen, and Alexis Feringa, 'Risk Management Guide forInformation Technology Systems', National Institute of Standards andTechnology, July, 2002

[15] ArashGhorbanniaDelavar, Mehdi ZekriyapanahGashti, Behroznori, Mohsen Nejadkheirallah, "RMSD: An optimal algorithm for distributed systems resource whit data mining mechanism", Canadian Journal on Artificial Intelligence, Machine Learning and Pattern Recognition Vol.2, No.2, February 2011